# Geo-Neutrinos

S.T. Dye[a,b]

[a]Department of Physics and Astronomy, University of Hawaii at Manoa,
2505 Correa Road, Honolulu, Hawaii, 96822 USA

[b]College of Natural Sciences, Hawaii Pacific University,
45-045 Kamehameha Highway, Kaneohe, Hawaii 96744 USA

This paper briefly reviews recent developments in the field of geo-neutrinos. It describes current and future detection projects, discusses modeling projects, suggests an observational program, and visits geo-reactor hypotheses.

## 1. INTRODUCTION

Geo-neutrinos are electron antineutrinos emitted in the beta decay of long-lived, terrestrial isotopes and their daughters. Due to the distributed nature of their source, geo-neutrinos are not suited for studies of neutrino oscillation. Geo-neutrino flux measurements do provide experimental evidence for the quantity and distribution of radioactive elements internally heating Earth. Radiogenic heating helps power plate tectonics, hot-spot volcanism, mantle convection, and possibly the geo-dynamo. Information on the extent and location of this heating better defines the thermal dynamics and chemical composition of Earth. Fiorentini et al. [1] provide a comprehensive review of geo-neutrinos.

Geo-neutrino detection typically uses the same technology employed by reactor antineutrino experiments for many decades [2]. In this technique, an array of hemispherical photomultiplier tubes monitors a large central volume of scintillating liquid. Free protons in the scintillating liquid are the targets for geo-neutrinos. The detected signal is a coincidence of products from the inverse beta interaction. A prompt positron provides a measure of the geo-neutrino energy, which is followed by a mono-energetic neutron capture. This technique allows a spectral measurement of geo-neutrinos from uranium-238 and thorium-232. Geo-neutrinos from all other isotopes lack the energy to initiate the inverse beta reaction on free protons. Fig. 1 shows the calculated geo-neutrino energy spectrum [3]. A project to directly measure these spectra is ongoing [4]. The highest energy geo-neutrinos derive only from uranium-238. This enables separate measurement of geo-neutrinos from uranium-238 and thorium-232. The traditional inverse beta coincidence technique affords scant information on geo-neutrino direction. This impedes determination of geo-neutrino source locations.

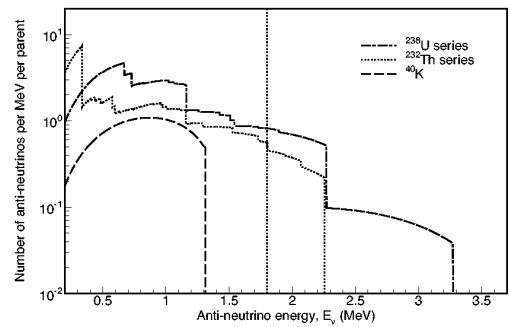

Figure 1. The calculated energy spectrum of geo-neutrinos from uranium-238, thorium-232, and potassium-40. Note the maximum energy of geo-neutrinos from potassium-40 is less than the threshold energy of the inverse beta reaction.

Two underground detectors are currently recording interactions of geo-neutrinos from uranium-238 and thorium-232, in Japan and Italy, using the inverse beta coincidence on free protons in scintillating liquid. Several other projects are in the planning stages. Future projects dedicated to measuring and modeling the planet's geo-neutrino flux would define the amount and distribution of heat producing elements in the Earth and provide transformative insights into the thermal history and dynamic processes of the mantle.

## 2. CURRENT PROJECTS

The Kamioka Liquid Scintillator Antineutrino Detector (KamLAND), operating since March 2002 in a mine in central Japan with 1000 tons of scintillating liquid, meets its physics agenda by measuring the oscillated spectrum of electron antineutrinos streaming from many nearby nuclear reactors. A recent analysis of KamLAND data estimates the geo-neutrino flux with 36% uncertainty when fixing the thorium to uranium mass ratio [5]. This estimate does not meaningfully constrain radiogenic heat production; it implies an upper limit greater than the measured terrestrial heat flow. Moreover, the measured spectrum shows no evidence for detection of the highest energy geo-neutrinos from uranium-238. The large antineutrino flux from the nearby reactors and radioactivity inside the detector compete with the geo-neutrino signal. However, removal of radioactivity in the scintillating liquid should be complete soon, with more sensitive geo-neutrino studies ensuing.

Borexino, operating since May 2007 with 300 tons of scintillating liquid in a tunnel in the Laboratori Nazionali del Gran Sasso in Italy, meets its physics agenda by measuring low energy solar neutrinos scattering on electrons [6]. An analysis of electron antineutrino data including geo-neutrinos is in progress. The geo-neutrino signal to background ratio is expected to be substantially larger for Borexino than for KamLAND due to lower radioactivity and lower flux of antineutrinos from nuclear reactors. Although their locations and sizes are not optimized for geo-neutrino investigations, KamLAND and Borexino are pioneering measurements that advance new scientific inquiry and aid future project development.

## 3. FUTURE PROJECTS

Scintillation detector projects in the design or planning stage offer opportunities for precision geo-neutrino measurements with low background. The next phase of the Sudbury Neutrino Observatory, called SNO+, would operate in a mine in Ontario, Canada, perhaps by early 2011 [7]. Comparable in size to KamLAND, it would be the world's deepest geo-neutrino observatory and the first situated mid-continent in North America. Other mid-continent projects would exploit the low reactor antineutrino flux at the Homestake mine in South Dakota [8] and at the Baksan Neutrino Observatory in the Caucasus [9]. The Low Energy Neutrino Astrophysics detector, called LENA, is under consideration for operation in a mine in Finland. It would be the largest project at 50,000 tons of scintillating liquid [10]. The Hawaii Anti-Neutrino Observatory, called Hanohano, is designed for deployment in the deep ocean with 10,000 tons of scintillating liquid. Operating in the tropical Pacific Ocean far from continental crust and nuclear reactors, it would principally observe geo-neutrinos from the mantle [11]. Being capable of redeployment at alternate sites it could potentially measure lateral heterogeneity of uranium and thorium in the mantle. The location and overburden of existing and proposed geo-neutrino project sites are compared in Table 1.

Table 1
Location and overburden of existing and proposed geo-neutrino project sites.

| Project | Lat (N) | Lon (E) | $H$ (mwe) |
|---|---|---|---|
| KamLAND | 36.43 | 137.31 | 2700 |
| Borexino | 42.45 | 13.57 | 3700 |
| SNO+ | 46.47 | -81.20 | 6000 |
| Homestake | 44.35 | -103.75 | 4200 |
| Baksan | 43.27 | 42.68 | 4800 |
| LENA | 63.66 | 26.05 | 4060 |
| Hanohano | 19.72 | -156.32 | 4500 |

## 4. MODELING PROJECTS

The two existing geo-neutrino flux models are largely the product of physicists [12,13]. These models enhance neutrino oscillation studies and enable geological investigations. Models typically establish a budget of uranium and thorium prescribed by a primitive mantle composition. Applying mass balance relationships to estimates of uranium and thorium in various Earth reservoirs predicts the distribution of these elements and the resulting geo-neutrino flux. The predictions by both models agree with the geo-neutrino flux recently estimated by KamLAND.

The predicted geo-neutrino flux at existing and proposed detection sites depends on the modeled distribution of uranium and thorium. Detection rates using the inverse beta reaction are given in terrestrial neutrino units (TNU). One TNU is equivalent to one interaction per $10^{32}$ free protons per year. The predicted detection rate depends strongly on location relative to the continents. This is illustrated in Fig. 2. The predicted contribution from a radial symmetric mantle ranges from 7-22 TNU.

## 5. OBSERVATIONAL PROGRAM

Determining the average concentrations of uranium and thorium in the mantle and continental crust is possible by comparing geo-neutrino observations at two geologically distinct locations. Whereas a continental observatory would primarily measure geo-neutrinos from the continental crust, an oceanic observatory would primarily measure geo-neutrinos from the mantle. Observations with Hanohano in the mid-Pacific and a detector with 3000 tons of scintillating liquid operating at Homestake would determine the global uranium content within about 20% uncertainty in three to four years [14]. The period of observation, which is greater for determination of thorium content, depends on the predicted geo-neutrino flux, background, and detection efficiencies. Complementary measurements of geo-neutrinos from mid-continental and mid-oceanic detectors would constrain the global content of uranium and thorium, and thereby radiogenic heating of the Earth.

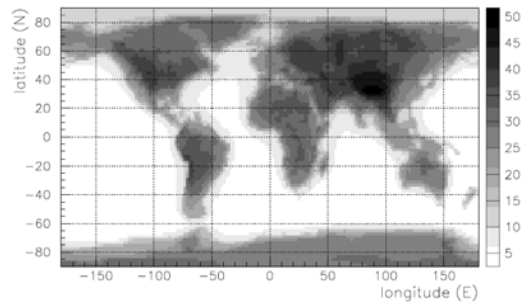

Figure 2. The geo-neutrino detection rate at the surface of the Earth due to only the crust in TNU is predicted using the methods and parameters from Enomoto et al. [13].

6. GEO-REACTOR

Whereas radiogenic heating of Earth is certain, although imperfectly quantified, heating by natural fission reactors is speculative. Provocatively, proposals suggest nuclear reactors may exist in or near Earth's core. These geo-reactors would emit electron antineutrinos just as do human-made nuclear reactors. KamLAND data restrict the power of an Earth-centered geo-reactor to be less than 20% of the terrestrial heat flow [5]. This eliminates some geo-reactor models [15] yet allows others; including one recently highlighted model that suggests a geo-reactor at the core-mantle boundary [16]. An antineutrino detector operating at a location where the flux from human-made reactors is a minimum, such as Hanohano in the mid-Pacific, would be sensitive to geo-reactors with power as low as a few percent of the terrestrial heat flow.

7. CONCLUSIONS

The detection of geo-neutrinos for the study of Earth's composition and energy dynamics is established by two ongoing projects. Meaningful constraints on terrestrial radiogenic heating are unlikely to result from these projects due to their limited statistical reach. Larger, strategically-located, future projects could estimate average uranium and thorium concentrations in the main Earth reservoirs and test geo-reactor hypotheses, thereby providing transformative insights into Earth's thermal history and dynamic processes. Measurement of geo-neutrino direction and determination of the flux due to potassium-40 would significantly advance these investigations but require new techniques.